\documentstyle[12pt]{article}
\begin{document}
{

\title{A new Proposal for a Quasielectron Trial Wavefunction
for the FQHE on a Disk}
\author{M. Kasner and W. Apel \\Physikalisch-Technische
Bundesanstalt
\\Postfach 3345
\\D--W--3300 Braunschweig, Germany}
\date{}
\maketitle

{\bf Abstract.}
In this letter, we propose a new quasielectron trial wavefunction for
$N$ interacting electrons in two dimensions moving in a strong magnetic field
in a
disk geometry. Requiring that the trial wavefunction
exhibits the correct filling factor
of a quasielectron wavefunction, we obtain $N+1$ angular momentum
eigenfunctions.
The expectation values of the energy are calculated
and compared with the data of an exact numerical diagonalization.

 \newpage

Trial wavefunctions have been used very successfully to describe the
low-energy properties of interacting electrons moving in two dimensions
in a strong magnetic field \cite{Prange}.
The prime example is the Laughlin wavefunction \cite{Lau1}, which describes
the ground state of the system at some of the filling factors of the lowest
spin-polarized Landau level (i.\ e.\  electronic densities), for which
one observes the
fractional quantum Hall effect (FQHE) \cite{TSG} .
Deviations from such a state to higher or lower filling factors are then
described by quasielectrons or quasiholes, for which trial wavefunctions
are also proposed \cite{Lau1}.
These are the constituents of the hierarchical picture \cite{Hal1,Halp}
which is used to describe FQHE states at the remaining rational
filling factors observed which are different from $1/m$ ($m$ odd integer).
The quasiparticles have also been proven to be a useful tool in numerical
studies of the FQHE system for small particle numbers.
Excitation energies have been calculated \cite{Halp,MoHal,HalRe,MacD}
and even some of the quasiparticle interactions have been obtained
\cite{Be,Endes}.
The situation is rather satisfying for the quasiholes, but, in the case of
the quasielectrons, questions remain.
Some of these we address in the present letter.
After introducing our notation, we present a new trial wavefunction for the
quasielectron state and then give as an illustration the first numerical
results for the expectation-value of the energy.\vspace*{5 mm}

We study the disk geometry in which the single particle basis of the lowest
Landau level is given by
\begin{equation}
\phi_{m}(z)=\frac{1}{\sqrt{2\pi2^{m}m!}}z^{m}e^{-\frac{1}{4}|z|^{2}} ,
\hspace{0.2cm} z=x-iy,
\end{equation}
$m=0,1,\ldots $ is the quantum number of the angular momentum,
 and the magnetic length $ l_{c}=\sqrt{\frac{\hbar}{|eB|}}$
is set to unity (the angular momentum $m$ is not to be confused with the
above odd integer $m$ of the filling factor).
For a description of a finite system confined to a disk, only
wavefunctions with
$0\le m\le m_{max}$ are kept in the basis, and $m_{max}$ then determines the
area $A$ via $A=2\pi (m_{max}+1)$. $A$ contains $m_{max}+1$ flux quanta.
We consider the case of $N$ spin-polarized electrons in the lowest Landau
level. Then, the definition of the filling factor, $\nu$, suited for
calculations with finite $N$, is $\nu=(N-1)/m_{max}$ \cite{Hal1}. When we
discuss, below, the filling factor of a wavefunction we shall determine it
from the maximum single particle angular momentum.
\\The trial wavefunction for the ground state at filling factor $1/m$
in the disk geometry \cite{Lau1} was proposed by Laughlin as
\begin{equation}
\Psi_{m}(z_{1},...,z_{N})=\prod^{N}_{i<j} (z_{i}-z_{j})^{m}e^{-\frac{1}{4}
\sum_{j=1}^{N}|z_{j}|^{2}} .
\end{equation}
Here, the maximum single particle angular momentum is $m_{max}=(N-1)m$, hence
$\nu=1/m$.
A \underline{quasihole} is defined as the ground state wavefunction
of a system the area of which is increased by
\underline{one additional flux quantum}. Its trial wavefunction is
\begin{equation}
\Psi^{(-,z_{0})}_{m}(z_{1},...,z_{N})=\prod_{j=1}^{N}(z_{j}-z_{0})\Psi_{m}.
\end{equation}
Similarly, a \underline{quasielectron} is the ground state of a system
containing  \underline{one flux } \newline
\underline{quantum less} than in the case of
$\nu = 1/m$.
The original proposal for its trial wavefunction reads
\begin{equation}
\Psi^{(+,z_{0}^{*})}_{m}(z_{1},...,z_{N})
=\prod^{N}_{j=1}(2\frac{\partial}{\partial z_{j}}
-z_{0}^{*})\Psi_{m},
\end{equation}
where the derivatives act only on the polynomial part of $\Psi_{m}$.
\vspace*{5 mm}

While it is not a priori obvious that $\Psi_{m}$, $\Psi_{m}^{(-,z_{0})}$ and
$\Psi_{m}^{(+,z_{0}^{*})}$ are good variational wavefunctions for the ground
state at these three different filling factors in the sense that they yield a
low energy-expectation-value for a given interaction, it is well known that
$\Psi_{m}$ and $\Psi_{m}^{(-,z_{0})}$ are exact eigenfunctions of a
Hamiltonian with short range interaction \cite{TruKiPoTa} for arbitrary values
of the parameter $z_{0}$ related to the position of the quasihole.

In order to get eigenfunctions of the total angular momentum M, suited for the
disk geometry, we expand the quasielectron trial wavefunction
$\Psi_{m}^{(+,z_{0}^{*})}$ with respect to $z_{0}^{*}$.
This yields $N+1$ linearly independent components (quasielectrons) whose
angular momenta $M$ vary between $M^{*}-N$ and $M^{*}$.
$M^{*}=(N-1)Nm/2$ is the total angular momentum of the ground
state wavefunction (2).
Now one clearly sees the problems with the proposal $\Psi_{m}^{(+,z_{0}^{*})}$
for the quasielectron.
The component with $M=M^{*}-1$ ($\propto (z_{0}^{*})^{N-1}$) is
\underline{zero}.
That means, one gets $N$, instead of $N+1$, quasielectrons.
This is different in the case of the quasiholes.
Second, all but one of the remaining $N$ quasielectrons have the wrong filling
factor $\nu = 1/m$, because their $m_{max}$ is $(N-1)m$.
These are the quasielectrons with $M=M^{*}-N+1, \dots, M ^{*}-2,M^{*}$ .
Thus, the whole wavefunction $\Psi_{m}^{(+,z_{0}^{*})}$ exhibits the wrong
filling factor $\nu = 1/m$, and is hence not suited for a quasielectron trial
wavefunction.
Only one component ($\propto (z_{0}^{*})^{0}$) has $m_{max}=(N-1)m-1$ and
thus shows the correct filling factor.
All these features of (4) limit the value of a comparison with exact numerical
calculations for small systems, cf.\ \cite{Kasner} and the Table below.

There are more proposals for quasielectron trial wavefunctions
\cite{MoHal,MacDGir,Jain}.
While these show the correct filling factor, they, unfortunately, give only one
component, i.\ e.\ $M=M^{*}-N$ (quasielectron at the origin).
Thus, apparently none of these trial wavefunctions in the disk geometry
corroborates the picture that one has $N+1$ quasielectrons becoming
degenerate in energy in the thermodynamic limit, just as electrons in a
Landau level, which is one of the main ingredients in the hierarchical
theory. \vspace*{7 mm}
\newpage
In order to improve this situation, we propose a different quasielectron
trial wavefunction
\begin{equation}
\prod_{j=1}^{N}[(N-1)m-z_{j}\partial_{z_{j}}]\Psi_{m}
\end{equation}
which we will motivate in the following.
The operator $z_{j}\partial_{z_{j}}$, acting only on the polynomial part,
yields the angular momentum of particle $j$.
Thus, the operation in (5) annihilates those parts of $\Psi_{m}$ in which the
single particle state with highest angular momentum $(N-1)m$ is occupied.
Hence,
$m_{max}=(N-1)m-1$ in (5) , and thus the filling factor comes out correctly
(it is \underline{larger} than $1/m$ by an amount of order $1/N$).
So far, we have only one state, i.\ e.\ the one at $M=M^{*}$.
In order to get the remaining $N$ functions, we apply a magnetic translation
followed by a projection to the correct filling factor.\vspace*{5 mm}

The result of the action of a magnetic translation
$T_{z_{0}}=e^{-i\vec{r}_{0} \vec{t}}$ ($\vec{t}=\vec{p}-e\vec{A}$)
on the basis (1) is
$T_{z_{0}}\phi_{m}(z)
=\frac{1}{\sqrt{2\pi 2^{m}m!}}
(z-z_{0})^{m}e^{-\frac{1}{4}(|z|^{2}+|z_{0}|^{2}-2zz_{0}^{*})}$, where
$z_{0}$ is given by the translation vector $\vec{r}_{0}$. The magnetic
translation operator for the $N$ particle system commutes with the
Hamiltonian in the thermodynamic limit.
If we now know an energetically favourable wavefunction (in the best case an
eigenfunction of $H$), we get more good candidates for low-energy
wavefunctions by applying a magnetic translation followed by a projection to
our $\nu$.
Starting out from (5), we construct in this way a set of quasielectrons
which should become degenerate in energy in the thermodynamic limit.
\vspace*{5 mm}

First, we demonstrate this translation and projection by applying it to
the Laughlin wavefunction (2).
Its polynomial part is unchanged, while additional terms $z_{j}z_{0}^{*}$
appear in the exponent.
We expand in $z_{0}^{*}$, in order to get angular momentum eigenfunctions.
Then, all coefficients of $(z_{0}^{*})^{n}, (n=1,2,\ldots)$ correspond to a
filling factor smaller than $1/m$.
The exception is the coefficient of $(z_{0}^{*})^{0}$, showing correctly
$\nu=1/m$.
Thus, projecting back to the original filling factor is equivalent to keeping
only the coefficient of $(z_{0}^{*})^{0}$ (putting $z_{0}^{*}=0$),
and it leads back to the function  (2).
(The term $e^{-\frac{1}{4}|z_{0}|^{2}}$ cancels out in the normalization.)
This supports the picture that the Laughlin wavefunction for the disk geometry
is non-degenerate in the thermodynamic limit. \vspace*{5 mm}

If we apply the same procedure to the component with $M=M^{*}+N$ ($z_{0}=0$)
of the quasihole wavefunction $\Psi_{m}^{(-,z_{0})}$, the magnetic translation
followed by a projection to $m_{max}=(N-1)m+1$ reproduces the function
$\Psi_{m}^{(-,z_{0})}$  completely.
This shows that the parameter $z_{0}$ in (3) is related to the translation
vector of a magnetic translation.
Hence, the (N+1) linearly independent quasihole wavefunctions, resulting from
an expansion of (3) in M-eigenfunctions, are degenerate in energy, already for
a finite system.
The physical meaning of the magnetic translation followed by the projection is
the following:
The position of the quasihole is shifted against the background of the Laughlin
wavefunction which is kept fixed \cite{Weller}. \vspace{5 mm}

In contrast to the case of $\Psi_{m}^{(-,z_{0})}$, however, Laughlin's
quasielectron function $\Psi_{m}^{(+,z_{0}^{*})}$ taken at $z_{0}^{*}=0$ is
left unchanged by a magnetic translation followed by the appropriate
projection.
Thus, in contrast to the case of the quasihole, this procedure does not lead
to a set of energetically degenerate quasielectron wavefunctions which, on the
other hand, is the fundamental assumption of the hierarchical model of the
FQHE. \vspace*{7 mm}

Applying now a magnetic translation followed by a projection to
$m_{max}=(N-1)m-1$ \hspace{0.1cm}(i.\ e.\
$z \rightarrow z-z_{0}, \hspace{0.1cm} \partial_{z} \rightarrow \partial_{z}$)
\hspace{0.1cm} to the function (5), we finally arrive at our new proposal for a
quasielectron trial wavefunction which reads
\begin{equation}
\Phi^{(+,z_{0})}_{m}(z_{1},\ldots ,z_{N})=\prod^{N}_{j=1}([(N-1)m-
z_{j}\partial_{z_{j}}]+z_{0}\partial_{z_{j}})\Psi_{m}.
\end{equation}

Expanding the product in powers of $z_{0}$, we  correctly find $(N+1)$
quasielectron states which are eigenstates of the total angular momentum
with the eigenvalues $M^{*}-N,...,M^{*}-1,M^{*}$. \vspace*{5 mm}

It is interesting to note that the state at angular momentum $M^{*}-N$ is equal
to the corresponding state of the proposal $\Psi_{m}^{(+,z_{0}^{*})}$,
but that all other states are different.
We note a further property of $\Phi_{m}^{(+,z_{0})}$, which makes it
appear a very natural choice.
The above mentioned discrepancies with the quasielectron trial wavefunction
do not occur in the spherical geometry, where the electrons move on the surface
of a sphere in a magnetic field produced by a magnetic monopole in the center
\cite{Hal1}.
In this geometry, the filling factor of the quasielectron trial wavefunction is
correct.
Also, there are $(N+1)$ such states.
Remarkably, $\Phi_{m}^{(+,z_{0})}$ is just the stereographic projection of the
corresponding quasielectron trial wavefunction on the sphere \cite{Hal1} onto
the plane.
In previous work \cite{FOC}, the stereographic projection has been employed
to relate the wavefunction $\Psi_{m}$ on the plane to the corresponding
wavefunction on the sphere. The relation between the quasielectron trial
wavefunction on the sphere and $\Phi_{m}^{(+,z_{0})}$ is a further
application of this mapping.\vspace*{5 mm}

We conclude by comparing the expectation-values for the energies of the
components of the
quasielectron $\Phi_{m}^{(+,z_{0})}$ with exact finite size numerical
diagonalizations in the disk geometry.
We choose a short range interaction \cite{Hal2}, where only the coefficient
corresponding to relative angular momentum one is kept in the expansion and
taken to be equal to its Coulomb value.
This interaction is viewed as giving the basic model which incorporates
all important features of the problem and from which one could approach
a realistic interaction in a perturbative way.

The energy-expectation-values of (6) were exactly calculated with an algebraic
program (MATHEMATICA).
The results are listed in the Table for $m=3$ and $N=6$ electrons, i.\ e.\
$m_{max}=14$.
The energy-expectation-values for the components of the quasielectron trial
wavefunction $\Phi_{m}^{(+,z_{0})}$ in the second column can be
compared with the
lowest eigenenergy of the exact numerical diagonalization in the third
column of the Table for each total angular momentum $M$.
We show the energy-expectation-values of the components of Laughlin's
quasielectron trial wavefunction in the fourth column of the Table, in spite
of the fact that a comparison with them is not justified because of the wrong
filling factor.
The only exception from this is the case $M=39$ where the two trial
wavefunctions
coincide.
For $M=40, 41, 42, 43,$ the energies of $\Phi_{m}^{(+,z_{0})}$
 are lower than those of the original proposal
$\Psi_{m}^{(+,z_{0}^{*})}$.
Since a particle in $\Psi_{m}^{(+,z_{0}^{*})}$  can occupy a state with angular
momentum 15
and a particle in $\Phi_{m}^{(+,z_{0})}$ can not, and thus particles
in $\Psi_{m}^{(+,z^{*}_{0})}$ can
lower their energy by keeping a larger mutual distance than those in
$\Phi_{m}^{(+,z_{0})}$,
this is all the more remarkable.
As mentioned above, a component of $\Psi_{m}^{(+,z_{0}^{*})}$ at $M=44$
does not
exist.
For $M=45$, the component of $\Psi_{m}^{(+,z_{0}^{*})}$ is obviously not
an allowed
variational wavefunction, it does not obey the constraint
(filling factor).
On the other hand, the energy of $\Phi_{m}^{(+,z_{0})}$ at $M=45$
is a good upper bound for the exact energy.

Viewed as a function of $M$, the expectation-values of the components of the
trial wavefunction $\Phi_{m}^{(+,z_{0})}$ follow the behavior of the
exact energies with the exception between $M=40$ and $M=41$;
the dependence of the variational energies
on $M$, however, is much stronger than that of the exact energies.
In any case, we do get an upper bound for the energy from
$\Phi_{m}^{(+,z_{0})}$.
While we expect from the above considerations that the energies of the
components of $\Phi_{m}^{(+,z_{0})}$ become degenerate in the thermodynamic
limit, we still find strong finite size corrections at $N=6$. \vspace*{5 mm}

In this letter, we have proposed a new quasielectron trial wavefunction
$\Phi_{m}^{(+,z_{0})}$ for the FQHE in the disk geometry.
The request for the correct filling factor is fulfilled.
By using the magnetic translation a whole set of quasielectron
wavefunctions for different total angular momenta has been constructed.
On the scale of the bandwidth, the magnitude of the energy is already for
$N=6$ quite reasonable compared with the energy resulting from the
exact diagonalization. An interesting, and, to our knowledge unique feature
of (6) is that the creation operator for the quasielectron depends
explicitely on the state it is applied to. Generalizing our approach to trial
wavefunctions with more quasielectrons we expect that the quasielectron
creation operator depends on the number of quasiparticles already present.
Work on the consequences for the statistics of these quasiparticles is under
way. \vspace*{5 mm}

We would like to thank Allan MacDonald for a stimulating discussion at
MPI Stuttgart.
We also appreciated very much the hospitality of our hosts at the
Max--Planck--Institut f\"ur
Festk\"orperforschung in Stuttgart.

\newpage
\noindent
\underline{Table} \\
\begin{tabular}[t]{|l|l|l||l|}
\hline
M & $\Phi_{m}^{(+,z_{0})}$  & Diagonalization & $\Psi_{m}^{(+,z_{0}^{*})}$ \\
\hline
39 & 0.2259 & 0.1839 & 0.2259 \\  \hline
40 & 0.2437 & 0.2168 & 0.2451 \\  \hline
41 & 0.3147 & 0.1995 & 0.3422 \\  \hline
42 & 0.3688 & 0.2025 & 0.4546 \\  \hline
43 & 0.4108 & 0.2078 & 0.5678 \\  \hline
44 & 0.4368 & 0.2145 & ---    \\  \hline
45 & 0.0846 & 0.0188 & 0      \\  \hline
\end{tabular} \\
\vspace{1.0cm} \\
The energy-expectation-values of the components of the
quasielectron wavefunction in a system with $N=6$ particles, $m=3$.
Column 1 is the total angular momentum $M$. Column 2 shows the data for our
trial wavefunction $\Phi_{m}^{(+,z_{0})}$, in column 3, the results of an exact
diagonalization are displayed. Column 4 shows the
data for Laughlin's quasielectron wavefunction
$\Psi_{m}^{(+,z_{0}^{*})}$.
}
\end{document}